\documentclass[aps,a4paper,pre,floatfix,showkeys,twocolumn]{revtex4}
\usepackage{graphicx}
\usepackage{amsmath}
\usepackage{amssymb}
\usepackage{wasysym}
\usepackage{setspace}
\usepackage{color}
\usepackage[dvipsnames]{xcolor}

\begin{document}

\title{Relevance of the Computational Models of Bacterial Interactions in the simulation of Biofilm Growth}

\author{Gabriel Santos-D\'{i}az$^{1}$}	

\author{\'Alvaro Rodr\'iguez-Rivas$^{3}$}

\author{Alejandro Cuetos$^{1,2}$}
\email{acuemen@upo.es}

\affiliation{$^1$Department of Physical, Chemical and Natural Systems, Universidad Pablo de Olavide, Sevilla, Spain}
\affiliation{$^2$ Center for Nanoscience and Sustainable Technologies (CNATS), Universidad Pablo de Olavide, Sevilla, Spain}

\affiliation{$^3$ Departamento de Matem\'atica Aplicada II, Universidad de Sevilla, E.T.S. de Ingenier\'{\i}a, C. de los Descubrimientos s/n. Pabell\'on Pza. de Am\'erica 41092 Sevilla (Spain)}

\begin{abstract}
This study explores the application of elongated particle interaction models, traditionally used in liquid crystal phase research, in the context of early bacterial biofilm development. Through computer simulations using an agent-based model, we have investigated the possibilities and limitations of modeling biofilm formation and growth using different models for interaction between bacteria, such as the Hertz model, Soft Repulsive Spherocylindrical (SRS) model, and attractive Kihara model. Our approach focuses on understanding how mechanical forces due to the interaction between cells, in addition to growth and diffusive parameters, influence the formation of complex bacterial communities. By comparing such force models, we evaluate their impact on the structural properties of bacterial microcolonies. The results indicate that, although the specific force model has some effect on biofilm properties, the intensity of the interaction between bacteria is the most important determinant. This study highlights the importance of properly selecting interaction strength in simulations to obtain realistic representations of biofilm growth, and suggests which adapted models of rod-shaped bacterial systems may offer a valid approach to study the dynamics of complex biofilms.
\end{abstract}




\maketitle

\section{Introduction}\label{sec1}

In theoretical and simulation studies, the way of modeling the interaction between the agents (molecules, colloidal particles, etc.) that constitute the system has always been recognized as a fundamental element. In the late 20th and early 21st centuries, the choice between the different models available was a major issue in the scientific literature. At that time, it was crucial to clarify whether the different options to choose from were able to realistically reproduce the complexity of experimental observations. This was especially relevant in the field of simulation and theoretical studies of non-spherical particle fluids, such as linear hydrocarbons, or prolate and oblate mesogens, where several models have been proposed as tools for the systematic study of complex fluids by means of computer simulation \cite{GB81,DEM91,RULL95,MWJ96,BOL97,KIH53,CUE05,MAR04C}. Following these ideas, we present in this work a systematic study of the choice of some of these models for the modeling and simulation of cell colonies. Specifically, in this work we focus on the study by means of computer simulation of the development of bacterial biofilms.

Bacterial biofilms are the most common form of bacterial communities. They consist of bacterial assemblages that, often starting from a single cell and through the process known as binary fission, can form complex structures of millions of individuals. To understand the formation and development of these communities several aspects must be taken into consideration. These include environmental factors such as nutrient and harmful chemical concentrations, the substrate on which the colony grows, or the presence of other bacterial species, among others \cite{COS95,FLE16}. In reaction to these conditions, cells exhibit various physiological responses that are typically genetically controlled and mediate their reactions to different environmental interactions.

Traditionally, the study of bacteria and biofilm responses to external stimuli has focused on the characteristics of the chemical environment in which cells grow. However, in recent years, there has been growing interest in the mechanical aspects of bacterial and biofilm responses \cite{DUF20,ISH20,PER15}. For instance, the movement of bacteria in bulk fluids, where hydrodynamic interactions with other cells and flow play a vital role \cite{ISH20,DUN13}, or the rheological properties of biofilms \cite{GEI22,MAR23,PAV15,PER15,ISH20} 

The mechanical response of individual bacteria and biofilms to applied forces and stresses is also an important factor to consider. For instance, there is ample evidence that growth-induced pressure results in a reduction in the rate of cell growth and reproduction in cellular ensembles. This effect has been observed in populations of microorganisms, such as yeast \cite{DEL16,ALR22} and bacteria \cite{STE89,VOL08,LAU19}, as well as in tissues \cite{SHR05,VAL22} and tumors \cite{HEL97,ALE13}. The ability of bacteria to respond to mechanical stimuli could be attributed to two key factors. Firstly, the presence of mechanotransduction systems in the bacterial wall facilitates the conversion of mechanical inputs into cellular responses \cite{PER17,GOR19}. Secondly, the interactions between cells and their environment, as well as between different cells. In this sense various types of forces have been identified at different stages of biofilm development \cite{EVE17,JIN20}. Surface forces, comprising electrostatic and van der Walls interactions, lubrication forces, or the so-called fimbrial forces \cite{APR11,REN11}. All these interactions can be repulsive or attractive.

In recent times, considerable effort has been focused on creating models and simulations to better understand the growth and development of biofilms, as well as other cellular communities such as tissues and tumors. Among the various approaches proposed, a subject on which it is possible to find very interesting reviews in the literature \cite{WAN10,DZ19,DEL22b}, Individual-Based Models (IbM) have gained considerable attraction and attention as an effective tool for studying these complex systems \cite{KRE01,DRA07,WAN10,GAR12,LIE15,ACE18,NIJ23}. This approach suggests that the process of growth in a cell community, from microcolonies to biofilms in the case of bacteria, can be effectively characterized by analyzing the key attributes of individual bacteria and the interactions between them. In IbM models it is possible to track and quantify the properties of each individual cell, which is one of their main advantages. In these models, the key ingredients are cell elongation and division, in some cases modulated by nutrient consumption, and an overdamped Brownian dynamics to simulate bacterial movement. It is in these dynamic equations where forces between cells and with other elements such as walls or obstacles must be introduced \cite{HOC10,EPS11}. In this context different mathematical models have been employed, aiming to capture the phenomenology of the forces between cells described above \cite{GAR12}. These models include linear springs \cite{DRA01}, or repulsive forces proportional to contact area \cite{STE95,PAL00}, to cite some early examples.

Possibly the most employed model is the derived from the Hertz theory of elastic contact \cite{LAN08}. This model has often been used to estimate the force between two bacteria in contact \cite{DOU20,YOU18,GRA14,GAR12}. The main limitation of this model in this context is that it has been proposed for spherical particles in contact, and therefore does not take into account the elongated shape of bacteria. Another limitation of the Hertz model is that, even with total overlap between the two cells, the force maintains a finite value. Although this is partly compensated for by the high value of Young's modulus found for bacteria \cite{TUS12}, the smoothness of the interaction makes it necessary to use very small time steps in computer simulation studies to avoid unrealistic overlaps between cells. The possibility of such spurious overlaps is more relevant in the simulation of compact colonies, where, as noted above, the self-induced stress can be very high.

As an alternative to the Hertz model, our group has recently proposed the use of models commonly used in other contexts, such as the study of elongated particle fluids. 
Thus, we have proposed the use of spherocylindrical models such as the Soft Repulsive Spherocylindrical (SRS) model \cite{EAR01,CUE02,CUE05,CUE17} or the attractive Kihara potential \cite{KIH53,CUE03} for modeling early biofilm development using IbM scenarios \cite{ACE18,LOB21,LOB21B,DEL22}. These models have the advantage of naturally introducing the spherocylindrical shape observed in rod-shaped bacteria. Moreover, these models rapidly become more repulsive as the overlap between interacting particles increases, thus imposing a constraint on this possibility. On the other hand, the main limitation for their use in the context of cellular simulation is that there is no physical reason to assume that these models reflect the functional form of steric interaction between two cells, beyond possible similarities with molecular or colloidal systems.

In any case, to the best of our knowledge, there are aspects of the use of possible cell-cell interaction models that previous studies have not clarified. For example, there is the basic question of whether the choice of one force model over another, beyond its physical realism, has a significant effect on the collective properties of cell colonies as observed by simulation. This would include testing the effect of the greater or lesser smoothness of the interaction, as well as the possible existence of attractive contact forces in addition to the repulsive forces due to steric interaction. To explore this issue, we have compared the properties of microcolonies generated by simulation when modeling bacteria interaction with the Hertz model, the SRS interaction potential or, in contrast to the repulsive nature of these two models, the attractive Kihara potential which additionally presents an attractive contribution. Our main objective is to elucidate the importance of choosing a particular cell-cell interaction model when proposing an agent-based algorithm for modeling biofilm development. We believe that this is an important aspect, and that it will be of considerable help to researchers interested in implementing models for the study of cellular communities. We also want to elucidate the effect of the strength of the interaction between bacteria, to determine the weight and relevance this parameter may have on the collective properties of the cell colonies. As in previous work \cite{ACE18,LOB21,DEL22}, we focus on the case of early microcolonies when they are still two-dimensional. We understand that our conclusions are valid and extendable to later situations, where colonies are more complex and three-dimensional.

\section{Methods}

\begin{figure*} [!t]
	\includegraphics[width =2\columnwidth]{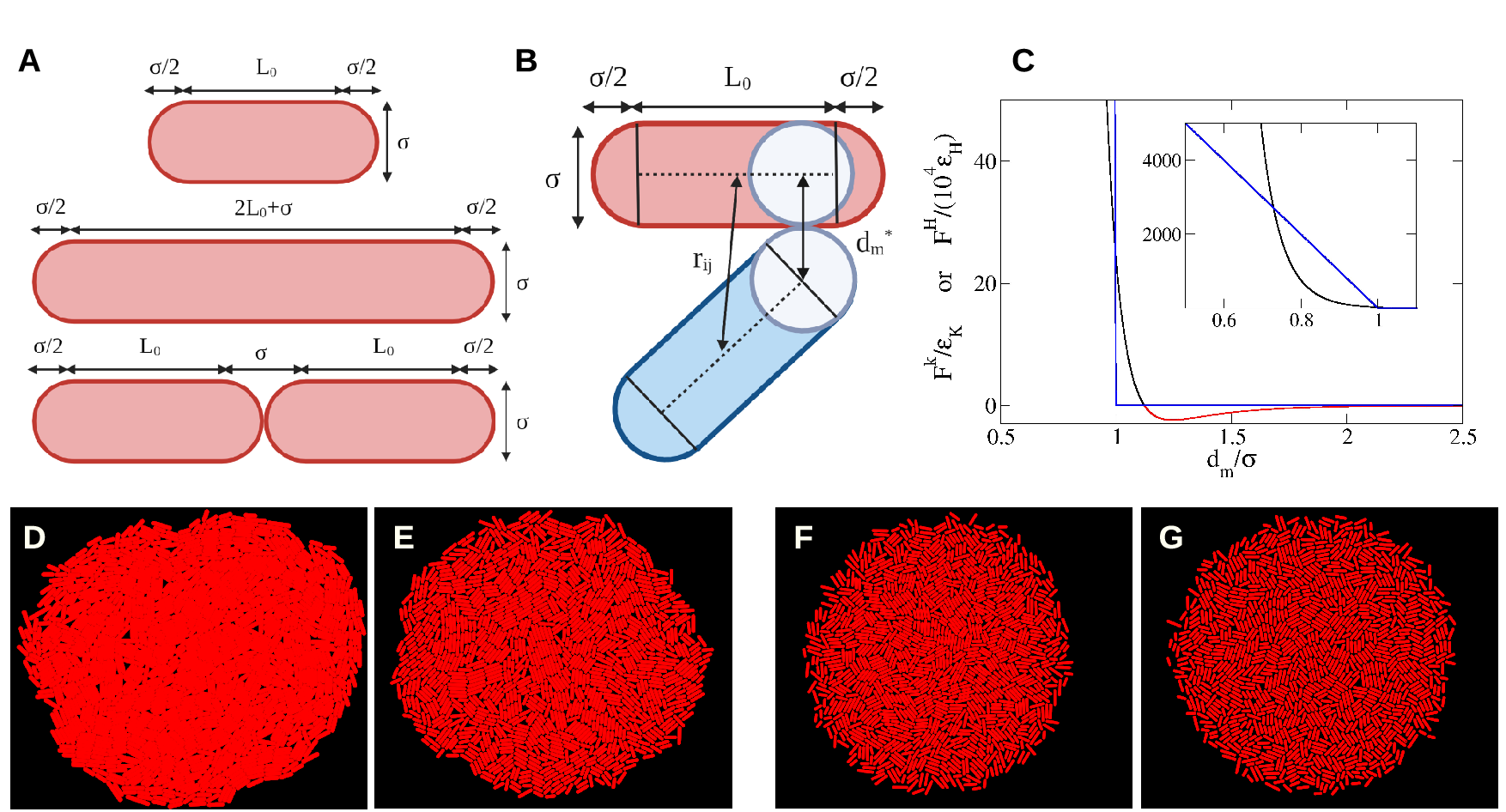}
	\caption{(A) Schematic representation of the geometrical parameters of the bacterial model. (B) Parameters involves in the calculation of the forces in SRS, attractive Kihara and Hertz models (see main text for details). (C) Comparison between the modulus of the forces calculated by SRS and attractive Kihara models ($F^K/\varepsilon_K$, black and red line respectively), and Hertz model ($F^H/10^4\varepsilon_H$, blue line) as function of the minimum distance between cells $d_m/\sigma$. The inset shows a zoom to the region of $d_m/\sigma \in [0.5,1.1]$. (D)-(G) Configurations of microcolonies with biomass $m(t)\approx 3500$ obtained with SRS model with $\varepsilon_K/k_BT=1$ (D and F) and  $\varepsilon_K/k_BT=100$ (E and G), at $\Gamma=10$ (D and E) and $\Gamma=0.1$ (F and G).}
	\label{figconf}
\end{figure*}

The simulation methodology employed in this work has many points in common with our previous work on biofilm simulation \cite{ACE18,LOB21,DEL22}. It should be noted that, as discussed in these previous studies, our model does not explicitly introduce relevant aspects such as the EPS matrix or the effect of water, these factors being considered implicit in the model parameters. We highlight here the most relevant details. As in previous studies, we have focused on the early stages of biofilm development, when microcolonies can be considered two-dimensional systems. In these stages, bacterial biofilms have the particularity that in the early stages of their development they are two-dimensional structures, evolving to three-dimensional systems in later stages of their development \cite{GRA14}. In addition, experimental setups have been described in which biofilms develop confined between two surfaces, always being two-dimensional \cite{VOL08}. Accordingly, rod-like bacteria have been modeled as the two-dimensional projection of the spherocylindrical particles. This kind of particle consists of a cylinder of instantaneous elongation $L$ capped by two hemispheres of diameter $\sigma$, the latter will be consider as the unit of length along this study. During the simulation, the diameter of the particles is going to remain constant, while the elongation of the cylinder will change over time. Thus, the elongation of the cylindrical part of a given particle grows exponentially, following the expression $L(t)=L^i \cdot exp(r^i\cdot(t-t_0))$. Here $r^i$ is the elongation rate of the bacterium $i$. $t_0$ is the instant when the last division occurred for bacterium $i$. $r^i$ is chosen at random from a Gaussian distribution centered at $r$ and relative standard deviation $s/r=0.1$. This exponential growth occurs from an initial elongation $L_0$. A time-dependent aspect ratio can then be defined for each bacterium as $L_i^*(t) =L_i(t)/\sigma + 1$. We will denote the initial aspect ratio as $L^*_0$. In this growth process, when a bacterium reaches an aspect ratio of $L^*_f =2L^*_0=2(L_0/\sigma+1)$ it splits into two identical particles, each with aspect ratio $L^*_0$ \cite{ACE18,DEL22}. The lengthening and division process is repeated from here for the two resulting cells. This process, as well as the main geometrical parameters describing the cells, are sketched in Fig.\ref{figconf}-A.

In addition to size, the position and orientation of the bacteria also evolves over time. This evolution has been simulated, as in previous work \cite{ACE18,LOB21,DEL22}, using Brownian dynamics, which incorporates both Brownian fluctuations in the position and orientation of the cells, as well as the motion caused by the interaction between the bacteria. Thus, the trajectory of the center of mass and orientation of its longitudinal axis of an individual bacterium $i$, defined by the vectors $\textbf{r}_i$ and $\textbf{u}_i$, evolves in the time according to the following set of equations:

\begin{equation}
\label{eq2b}
\begin{split}
{\bf r}_i^{||}(t+\Delta t) =
{\bf r}_i^{||}(t)+\frac{D_{i||}}{k_BT} {\bf F}_i^{||}(t)\Delta t + \\
\,\,\,\,\,\,\,\,\,\, + (2D_{i||} \Delta t)^{1/2} R^{||} {\bf \hat{u}}_i(t)
\end{split}
\end{equation}

\begin{equation}
\label{eq3b}
\begin{split}
{\bf r}_i^{\perp}(t+\Delta t) =    {\bf r}_i^{\perp}(t)+
\frac{D_{i\perp}}{k_BT} {\bf F}_i^{\perp}(t)\Delta
t+ \\
\,\,\,+ (2D_{i\perp} \Delta t)^{1/2} R^{\perp} \textbf{\^{v}}_i(t) \\
\end{split}
\end{equation}

\begin{equation}
\label{eq4b}
\begin{split}
\textbf{\^{u}}_i(t+\Delta t) = \textbf{\^{u}}_i(t)+
\frac{D_{i\vartheta}}{k_BT} {\bf T}_i(t)\times
\textbf{\^{u}}_i(t)\Delta t+ \\
\,\,\,+ (2D_{i\vartheta} \Delta t)^{1/2} R^{\vartheta}
\textbf{\^{v}}_i(t)
\end{split}
\end{equation}

\noindent being $\textbf{r}_i^{\parallel}$ and $\textbf{r}_i^{\perp}$ the projections of $\textbf{r}_i$ on the directions parallel and perpendicular to $\textbf{\^{u}}_i$, respectively. $\textbf{F}_i^{\parallel}$ and $\textbf{F}_i^{\perp}$ are the parallel and perpendicular components of the total force acting on $i$ and ${\bf T}_i$ is the total torque due to the interactions with other particles of the fluid \cite{VEG90}. The Brownian dynamic is induced through a set of independent Gaussian random numbers of variance 1 and zero mean: $R^{\parallel}$, $R^{\perp}$ and $R^{\vartheta}$. $\textbf{\^{v}}_{i}$ is an unitary vector perpendicular to ${\textbf{\^{u}}}_i$. In this equations $k_B$ is the Boltzmann constant, and $T$ the temperature. $\Delta t$ is the time step. 

For the diffusion coefficients, $D_{i\parallel}$, $D_{i\perp}$ and $D_{i\vartheta}$, as in previous works \cite{DEL22} we have employed expressions calculated by a method similar to that proposed by Bonet Avalos and co-workers \cite{BON94}. These diffusion coefficients depend on the size of the particles and they must be calculated for each bacterium at each time step. Taking into account this variation in size, and to gain computational efficiency, we have adjusted the results of the method proposed in \cite{BON94} to the following expressions:

\begin{eqnarray}
D_{\parallel}/D_0 &=&-0.0198\cdot ln(L^*)+ 0.0777+\frac{0.0437}{L^*}\nonumber\\
&&-\frac{0.0158}{L^{*2}}\nonumber\\
D_{\perp}/D_0 &=& -0.0119\cdot ln(L^*) + 0.0452 + \frac{0.0796}{L^*}\nonumber\\
&&-\frac{0.0190}{L^{*2}}\\
D_{\vartheta}\,\sigma^2/D_0 &=& -0.0002\cdot ln(L^*)+0.0012 - \frac{0.0243}{L^*}\nonumber\\
&&+\frac{0.3233}{L^{*2}}
+\frac{0.2597}{L^{*3}}-\frac{0.0483}{L^{*4}}\nonumber
\end{eqnarray}

\noindent depending on the diffusional parameter $D_0=D_ 0^*\sigma^2/\tau$, with $\tau$ the time unit. This diffusional parameter accounts for all aspects that affect the diffusion of an isolated bacterium, such as the effect of the solvent, friction with the substrate  or the adhesion to the substrate.

An important aspect of our work is how to model the forces between bacteria. As mentioned in the introduction, one of the aims of this study is to compare the effects of using one or the other model to represent the forces. As discussed there, in this paper we have chosen to compare two of the models that have been employed in the past in modeling the interaction between cells. On the one hand, SRS and attractive Kihara models, with and without attractive part. On the other hand, the Hertz elastic force model. We proceed to present these models in detail.

The Kihara interaction potential \cite{KIH53,CUE03} is a model used in the past to simulate the liquid crystal phase diagram of rod-like particles. The mathematical expression of this potential is:

\begin{equation}\label{eq1}
U_{ij}\,= \left\{ \begin{array}{cc}
4 \varepsilon_{K} \left[ \left(\frac{1}{d^*_m}\right)^{12}-\left(\frac{1}{d^*_m}\right)^{6}- E_{cu} \right] & ~~   d_{m} \leq d_{cu}
\\
0 & ~~ d_{m} > d_{cu} \end{array} \right.
\end{equation}

\noindent where $i$ and $j$ are generic particles (bacteria).  $d_m^*=d_m/\sigma$ is the shorter distance between them, calculated as the minimum distance between two spherocylinders (see Fig.\,\ref{figconf}-B) \cite{VEG94}. $d_{cu}$ is the distance at what the interaction is truncated and shifted, 
and $E_{cu} =d_{cu}^{*-12} - d_{cu}^{*-6}$. If $d_{cu} \le \sqrt[6]{2}\sigma$ the interaction is purely repulsive, while if $d_{cu}$ is bigger than this value the interaction is attractive for interparticle distances bigger than $\sqrt[6]{2}\sigma$. $\varepsilon_K$ represents the strength of the interaction in the SRS and attractive Kihara models. In this study $d_{cu}$ has been set to $\sqrt[6]{2}\sigma$ for the SRS model, and $2.5\sigma$ for the attractive Kihara models. In other words, the SRS potential is a purely repulsive version of Kihara's model. From this expression of the interaction potential is easy to obtain the forces and torques required in equations \ref{eq2b} to \ref{eq4b} \cite{VEG90}. Specifically, the explicit expression for the force that particle $i$ exerts on particle $j$ is

\begin{equation}\label{eq2}
{\bf F^K}_{ij}=\left\{ \begin{array}{cc}
-\varepsilon_K \left[\frac{24}{d^{*7}_m} - \frac{48}{d^{*13}_m} \right]\cdot {\bf \hat{e}_{ij}} & ~~   d^*_{m} \leq d^*_{cu}
\\
0 & ~~ d^*_{m} > d^*_{cu} \end{array} \right.
\end{equation}

In this equation ${\bf \hat{e}_{ij}}$ is the unit vector in the direction of the shortest distance from particle $i$ to $j$.

The Hertzian elastic model can be write as \cite{YOU18}

\begin{equation}\label{eq3}
{\bf F^H}_{ij}=\left\{ \begin{array}{cc}
\varepsilon_H \left( 1-d^*_m  \right)^{3/2}\cdot {\bf \hat{e}_{ij}}  & ~~   d^*_{m} \leq 1
\\
0 & ~~ d^*_{m} > 1 \end{array} \right.  
\end{equation}

\noindent with $\varepsilon_H$ the strength of the Hertzian interaction, depending on the bacteria Young's modulus and size. This force is repulsive whenever there is an overlap between two bacteria, and zero in other cases. Figure \ref{figconf}-B shows the geometrical parameters involved in both models. Fig.\ref{figconf}--C shows the dependence of the different cell-cell interaction models on the minimum distance between cells. Comparing now the repulsive models of SRS and Hertz, Fig.\ref{figconf}--C shows how at short overlaps between particles both models exert a similar repulsion (for equivalent values of the force intensity). At larger overlaps, the Hertz model is smoother, showing less repulsion than the SRS model. 

As shown in our previous studies  \cite{ACE18,DEL22}, the properties of microcolonies are highly dependent on the so-called $\Gamma$ parameter, which summarizes the competition between bacterial growth and bacterial diffusion. This parameter is defined as follows

\begin{equation}\label{eq4}
\Gamma =\frac{t_{dif}}{t_{gr}}
\end{equation}

Here $t_{dif}$ is the average time required by an isolated particle of constant aspect ratio $L^*_0$ to diffuse a distance $\sigma$ by Brownian diffusion, and $t_{gr}$ the time need by an average bacterium to reach the aspect ratio $L^*_f$ from its initial aspect ratio. $t_{dif}$ is dependent through $D_0$ on all the factors that affect the mobility of an isolated particle, such as the viscosity of the medium, the friction with the substrate or the adhesion of the bacteria to the substrate. In this work, $t_{gr}$ is a characteristic of the cells themselves, not conditioned in our model by environmental factors such as nutrient concentration or pressure felt by the cells. For a given value of $L^*_0$, $\Gamma$ is depending both on the diffusional parameter $D_0$ and on growth rate $r$. In the next section we will discuss the role of possible types of interaction between bacteria in different regimes characterized by different values of $\Gamma$. In each of these cases, to motorize the evolution of the microcolony we have estimated de amount of biomass in the microcolony, $m(t)$, defined as

\begin{equation}\label{eq5}
m(t)= \sum_{i=1}^{N(t)} L^*_i(t)
\end{equation}

\noindent being $N(t)$ the number of cells at time $t$. m(t) is time-dependent because $N(t)$ and $L^*_i(t)$ vary over time.

To determine the shape and size of the microcolony, we have obtained the ellipse that best fits the distribution of particles. For this we have calculated the components of inertia tensor as
\begin{equation}\label{eq5b}
\displaystyle{I_{\alpha,\beta}=\frac{1}{N(t)} \sum_{i=1}^{N(t)} \left(\delta_{\alpha,\beta} (\sum_{k=\alpha,\beta} r^k_{i})-  r^{\alpha}_{i} r^{\beta}_{i}\right)}
\end{equation}

Here $\alpha$ and $\beta$ indicate the coordinates $x$ or $y$, $\delta_{\alpha,\beta}$ is the Kronecker delta and $r^{\alpha}_i$ is the corresponding coordinate of the vector from the center of mass of the microcolony to the position of the bacterium $i$. Diagonalizing this tensor is possible to calculate the two semi-axes, $a>b$, of the ellipse that best fit the distribution of bacteria in the microcolony \cite{KAR07}. From this calculation, the area covered by the micro-colony can be estimated as the area of these ellipses $A_e=\pi ab$. Therefore, the global density of the microcolony is calculated as $\rho(t)=m(t)/A_e(t)$. Moreover, the shape of the microcolony can be measured by the eccentricity. The square of this eccentricity is defined as follows:

\begin{equation}\label{eq6}
e^2(t) = 1-\frac{b^2}{a^2}
\end{equation}

\noindent which approaches $0$ for  circular microcolonies. Additionally, to measure the orientational correlation of the particles the nematic order parameter $S_2(t)$ has been calculated by the standard procedure of diagonalizing a  traceless symmetric tensor build with the orientation vectors of all the particles \cite{DEL22,ALL93,MER92}. This nematic order parameter is 1 if all particles are parallel, and 0 if they are randomly oriented.

The value of $\rho$ defined above provides global information on the compactness of the microcolony. To obtain information on the differences in the packing of bacteria in different areas of the biofilm, we have used the coverage profile $g_c(r_{cm})$. This function is defined as the fraction of the surface covered by bacteria at a distance $r_{cm}$ from the microcolony center of mass. In the calculation of this function, a large number of random points are generated at a distance $r+dr$ from the center of mass of the microcolony, evaluating $g_c(r_{cm})$ as the fraction of these points that fall in the area occupied by a bacterium. The behavior of this function obtained with our simulation algorithm was compared with experimental results in a previous work \cite{ACE18}, finding good agreement.

As we have commented, the main goal of this work is to study how the characteristic of the option to model the interaction forces between bacteria influences the structural properties of the microcolony at different values of $\Gamma$. From our previous work \cite{DEL22}, the different regimes in biofilm evolution as a function of $\Gamma$ values are known. Taking this previous information into account, we have selected as significant values of the whole range $\Gamma=10, 1$ and $0.1$. To obtain these values, the growth rate has been chosen as $r\cdot\tau=0.26, 0.026$ and $0.0027$, respectively, while keeping $D^*_0=0.1$ in all cases. Specifying now for the interaction models used, for the SRS model ($d_{cu} = \sqrt[6]{2}\sigma$) and the attractive Kihara model ($d_{cu} = 2.5\sigma$), force intensities have been chosen in the interval $\varepsilon_K/k_B T\in[1,100]$. In the case of the Hertz model, the force intensity changes in the interval $\varepsilon_H/(10^4k_B T)\in[1,100]$. These intervals in the strength of the potentials have been chosen to explore a wide range of values, up to the saturation of the properties explored.

Throughout the simulations, the time step $\Delta t$ has been limited to values less than $10^{-3}\tau$, while ensuring that at each step the displacement of the center of mass of each cell was less than $0.01\sigma$. For all the cases analyzed in this work and in order to obtain statistical averages of the observables of interest, $500$ independent trajectories were run, all starting with a single particle of aspect ratio $L^*_0$ at the center of the simulation box.

\section{Results}

Biofilms are compact systems in which bacteria exert forces on each other. In real systems, this can lead to deformation of the cell walls. As the IbM models considered in this study are intrinsically rigid, this deformation is not considered. The equivalent consequence of the forces between bacteria in our algorithms would be the overlap  between particles. 

This overlap would result in distances between bacteria smaller than the cell diameter. These overlaps, which would be the equivalent of deformation in flexible models, will be unrealistic if they exceed a certain threshold. Therefore, a first indication of the realism, or the limitations, of each of the models at the different interaction intensities, would be to estimate the average overlap between particles. To quantify the magnitude of this overlap we have calculated the radial distribution function, $g(r)$. This distribution function gives information on how the structure around an average particle changes. This function is calculated as $g(r) \propto \left\langle  \delta(r-|r_i-r_j|) \right\rangle$ \cite{ALL93}, with the average being over all pairs of particles $i$ and $j$. The position of the first peak of this function, $r_1$, gives information about the average distance between nearest neighbors, thus obtaining information about the overlap between cells.

\begin{figure}[!t]
	\center
	\includegraphics[width =\columnwidth]{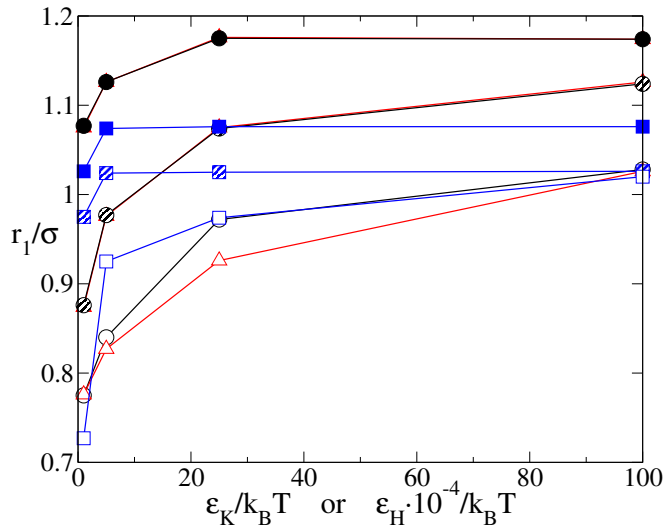}
	\caption{Position of the first peak of the distribution function in units of bacterial diameter $r_1/\sigma$, for microcolonies with biomass $m(t)=2000$ obtained with the SRS model (black lines and circles), the attractive Kihara model (red lines and triangles) and the Hertz model (blue lines and squares) as a function of the interaction strength ($\varepsilon_{K}/k_BT$ for the SRS and attractive Kihara models and $\varepsilon_H\cdot10^{-4}/k_BT$ for Hertz). The cases with $\Gamma=10$ (open symbols), $1$ (dashed symbols) and $0.1$ (solid symbols) are shown.}
	\label{fig1}
\end{figure}

Fig.\,\ref{fig1} shows the distance between first neighbors, $r_1$, for all the models of force considered in this work as a function of the strength of the interaction  and at the three values of $\Gamma$ explored. All the values shown in this figure are obtained averaging over 500 independent configurations with value of biomass $m(t)=2000$. As a first result for all models, the larger the $\Gamma$ the smaller the average distance between the first neighbors. This is to be expected, since the higher the $\Gamma$ the more compact the microcolony is \cite{ACE18,DEL22},  and therefore the closer together the particles are on average.

Focusing now in the case of $\Gamma=10$, Fig.\,\ref{fig1} shows as for low values of the strength of the interaction there is a high level of overlap between particles. This is dramatic in the case of the Hertz model with $\varepsilon_H = 10^4k_B T$, for which $r_1$ is less than $0.75$, indicating an average overlap of $25$ percent of the cell radius between nearest neighbors. Both the attractive Kihara and SRS models also show a significant level of overlap between particles, with $r_1$ less than $0.8$ for $\varepsilon_K=1k_BT$.

In all models, as the strength of the interaction grows, the distance between the first neighbors also increases, notably the rapid increment of $r_1$ for the Hertz model. For stronger interactions there is a saturation in the value of $r_1$, and the distance between first neighbors tends to a value independent of the intensity of the force. Thus, for example, in SRS and attractive Kihara models the first-neighbor distance changes little when $\varepsilon_K$ changes from $25k_BT$ to $100k_BT$, as it does in Hertz's model when $\varepsilon_H$ goes from $25\cdot10^4k_BT$ to $100\cdot10^4k_BT$. This confirms that this magnitude is quite independent at high values of the interaction intensity, maintaining the integrity of the cell membrane, and especially for the Hertz model. Interestingly, at intermediate values of the interaction strength, the attractive Kihara model shows a smaller distance between the closest neighbors $r_1$ than the other two repulsive models. This is a consequence of increased aggregation due to the attraction between cells. These differences between repulsive and attractive models disappear at higher values of the interaction strength. Thus, at the highest values explored, and for $\Gamma=10$, $r_1$ tends to $1.01$ in all cases.  This value indicates a small average overlap between bacteria, which is to be expected at this value of $\Gamma$ where colonies tend to be very compact \cite{ACE18,DEL22}. 

Focusing now on the case of lower values of $\Gamma$, the dependence of $r_1$ on the interaction intensity shows a qualitative behavior similar to that described above, although with smaller values of $r_1$. This indicates a low average overlap between cells, which is to be expected due to the lower compactness of the colonies. It is remarkable the coincidence at low values of $\Gamma$ ($\Gamma = 1$ and $0.1$) of the values of $r_1$ obtained for attractive Kihara and SRS models. This seems to indicate that, due to the higher relevance of bacterial diffusion at these values of $\Gamma$, possible attractive cell-cell interactions do not play a significant role. We will later qualify this conclusion by showing that at lower values of $m(t)$ the attractive character of the interaction does introduce differences. As in the case of $\Gamma=10$, $r_1$ is almost independent of the interaction strength for intermediate and high values of $\varepsilon_{K}$ and $\varepsilon_H$. In this range, the differences observed in Fig. \ref{fig1} can be explained by the value of $\Gamma$ (higher dispersion of cells for smaller $\Gamma$), and by the characteristics of the interaction model, with a longer repulsive tail for the SRS models (see equations \ref{eq2} and \ref{eq3}). 

In conclusion, it can be seen from Fig.\,\ref{fig1} that small values of the intensity of the interaction between bacteria lead to unrealistic situations, in which very high overlaps between bacteria occur. These overlaps could be observed, for instance, in the configuration obtained with the SRS model with $\varepsilon_{K}=1k_BT$ for $\Gamma=10$ at high value of the biomass (Fig.\,\ref{figconf}-D). In contrast, for this model and at this value of $\Gamma$ but at higher strength of the interaction ($\varepsilon_{K}=100k_BT$, Fig.\,\ref{figconf}-E) the bacteria show greater separation from each other, being the lower crowding in this case apparent by visual inspection. Observing the configurations at low value of $\Gamma$ ($\Gamma=0.1$, with $\varepsilon_{K}=1k_BT$ and $100k_BT$ in panels F and G of Fig.\,\ref{figconf}, respectively), it is not appreciate relevant different between then, confirmed the conclusions obtained from the analysis of position of first neighbors $r_1$ discussed up to now. It should be noted that these overlaps are not caused by the existence of attractive forces (there is an almost absolute coincidence between the attractive Kihara and SRS models at this value of $m(t)$), but by the internal pressures in the microcolony caused by the growth of the cells. Thus, the overlap is greater at higher values of $\Gamma$, where the colonies are more compact. In contrast, once a certain level of force strength is exceeded, this parameter has very little influence on the distance between first neighbors and on the overlap between near neighbors.

\begin{figure}[!t]
	\center
	\includegraphics[width =\columnwidth]{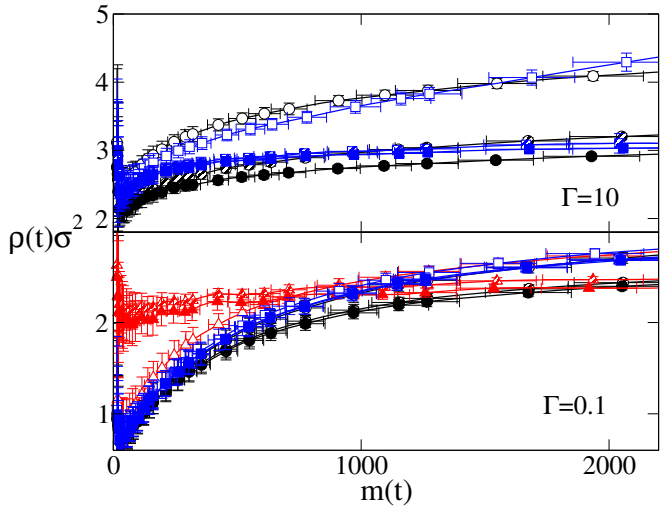}
	\caption{Cell density in reduced units $\rho\sigma^2$ as function of the biomass $m(t)$ for $\Gamma=10$ (top panel) and $\Gamma=0.1$ (bottom panel). In both panels are represented results obtained with SRS model (black lines and circles), attractive Kihara (red lines and triangles) and Hertz model (blue lines and squares). Open, dashed and solid symbols are for $\varepsilon_K= k_BT, 25k_BT$ and $100k_BT$ respectively in the case of SRS and attractive Kihara models, and for  $\varepsilon_H= 10^4k_BT, 25\cdot10^4k_BT$ and $100\cdot10^4k_BT$ in the case of Hertz model.}
	\label{fig2}
\end{figure}

This increase in overlaps between particles at low value of strength intensity has relevant effects on some macroscopic properties. For example, the impact of overlap between particles is clear in colony density ($\rho\sigma^2$ in reduced units, Figure \ref{fig2}), where higher overlap values translate into denser colonies. Fig.\,\ref{fig2} shows that this change is more relevant at high values of $\Gamma$. Thus, for $\Gamma=10$ (top panel of Fig.\,\ref{fig2}), the density $\rho\sigma^2$ in the case of the SRS model with $\varepsilon_K=1k_BT$ is much larger than for the other values of $\varepsilon_K$ shown ($25k_BT$ and $100k_BT$). A similar trend is observed in the case of the Hertz model for the density obtained with the lower strength forces explored ($\varepsilon_H=10^4 k_BT$) compared to the higher strength forces ($\varepsilon_H=25\cdot 10^4 k_BT$ and $100\cdot10^4 k_BT$), with values of the density similar to the obtained for SRS model. At this value of $\Gamma$ the dependence of density on biomass in the case of the attractive Kihara model is coincident over the entire biomass range for a given value of $\varepsilon_K$ with that described for the SRS model. For this reason, data for this case have not been plotted in top panel of Fig.\,\ref{fig2}. Thus, for $\Gamma =10$, the attractive and SRS potentials have the same density values over the entire biomass range for a given value of $\varepsilon_K$. This indicates that, in compact colonies due to the competition between diffusion and growth \cite{ACE18,DEL22}, the attractiveness of the interaction does not play a relevant role. 

At intermediate and low values of $\Gamma$ ($\Gamma=1$ not shown, and $\Gamma=0.1$, bottom panel of Fig.\,\ref{fig2}) the situation is very similar, but with some relevant qualitative differences. In any case, the bottom panel of Fig.\,\ref{fig2} shows that the changes of the density values for the different potentials and forces studied are smaller than for $\Gamma=10$. Within this smaller variation of the results, Fig.\,\ref{fig2} indicates that for this value of $\Gamma$ the trend is the same as the one mentioned for $\Gamma=10$, so that the potentials presenting a higher overlap, measured with the value of $r_1$ shown in Fig.\,\ref{figconf}, induce the formation of microcolonies of higher density.  

Despite the aforementioned, at this low value of $\Gamma$ an important qualitative difference is observed in the case of attractive Kihara potential. Thus, when the strength in this model is sufficiently intense ($\varepsilon_K=25k_BT$ and $100k_BT$), at low values of biomass a different behavior is observed than in the repulsive models. Whereas for low and intermediate values of biomass, $\rho\sigma^2$ shows higher values for the attractive models with high interactions than for the SRS or Hertz potentials. At higher values of $m(t)$ and in this range of force intensity, the density in all models is similar, something expected from the coincidence of the value of $r_1$ among these models for $m(t)=2000$ discussed in Fig.\,\ref{fig1}. This phenomenology is indicating that, although at this low value of $\Gamma$ bacterial diffusion dominates over growth and cells tend to disperse on the surface, if attractive interactions are sufficiently intense, dispersal is slowed by attraction between cells, which favour their staying together. Thus, although attractive interactions do not seem to play a relevant role when the colony is compact, it is relevant in situations where without these attractive interactions cells would trend to disperse.

\begin{figure*}[!th]
	\center
	\includegraphics[width =2\columnwidth]{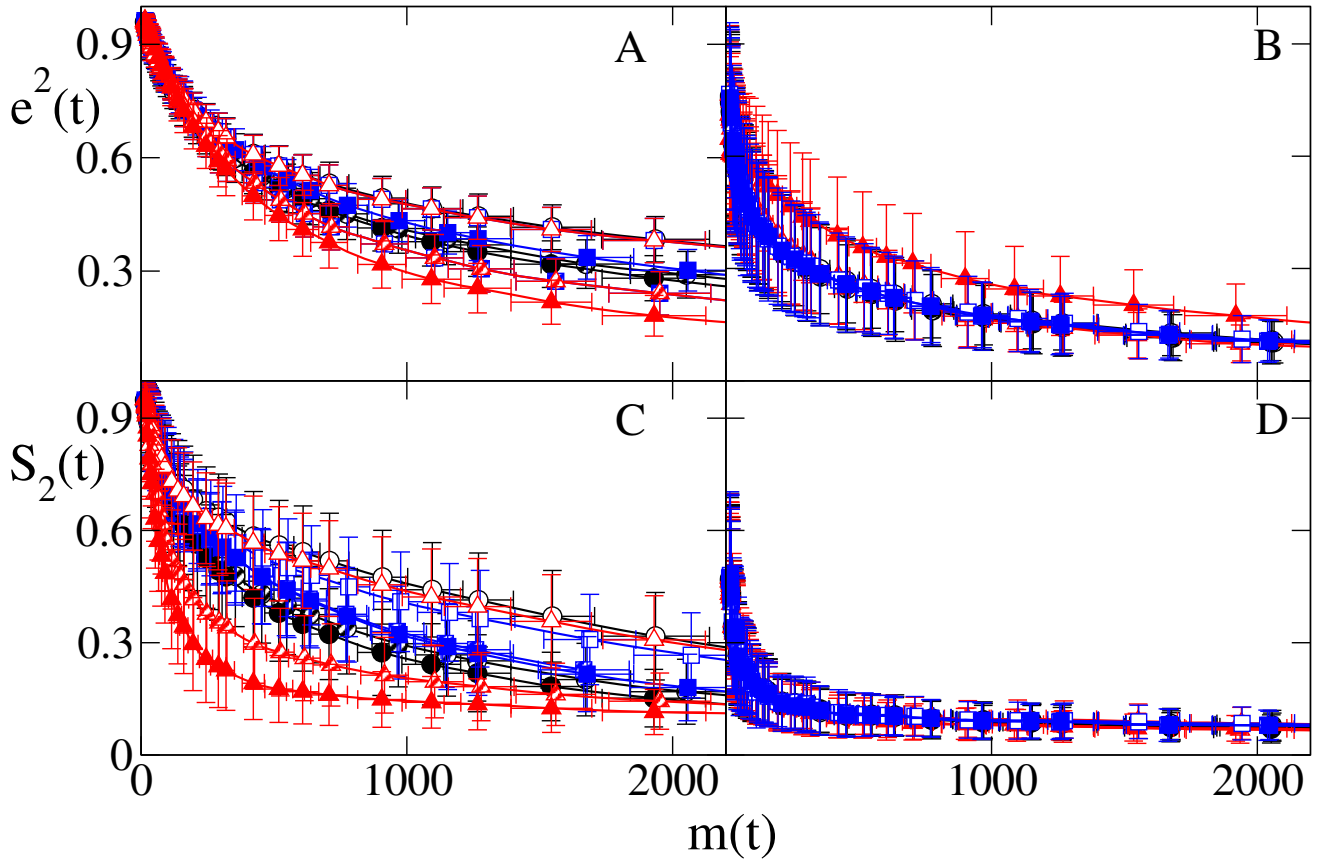}
	\caption{Square of the eccentricity of the microcolony ($e^2(t)$) and nematic order parameter ($S_2(t)$) as a function of the biomass $m(t)$, for $\Gamma=10$ (A) and (C), and for $\Gamma=0.1$ (B) and (D). The results obtained with  SRS, attractive Kihara and Hertz models are represent by the same symbols than in Fig.\,\ref{fig2}. 
	}
	\label{fig2b}
\end{figure*}

In summary, the results discussed up to now confirm the general behavior obtained in previous studies \cite{ACE18,DEL22}, which indicated that denser microcolonies are formed at higher $\Gamma$ values. But we see that this density depend also on the model and, above all, the strength of the interaction between cells. Thus, low interactions result in high densities, as overlapping between bacteria becomes possible. Conversely, more intense repulsive interactions between bacteria result in lower density values. This has consequences for the value of other macroscopic quantities that depend on density. For example, previous studies \cite{ACE18,DEL22} shown that higher density values result in more ellipsoidal colonies (higher eccentricity values), and with higher orientational order (higher value of the nematic order parameter). 

Fig.\,\ref{fig2b} confirms this dependence in the case of $\Gamma=10$. In this case (panels A and C) it can be seen that the lower the force intensity, and therefore the cell density, the higher the eccentricity and the nematic order parameter for the different models. In any case it is not a very strong dependence, and with all intensities the same conclusions would be reached as in previous work. At low values of $\Gamma$ (0.1 in panels B and D for $e^2(t)$ and $S_2(t)$) the results are very independent of the force model and intensity, confirming that in non-compact colonies this is not a relevant factor.

\begin{figure*}[!th]
	\center
	\includegraphics[width =2\columnwidth]{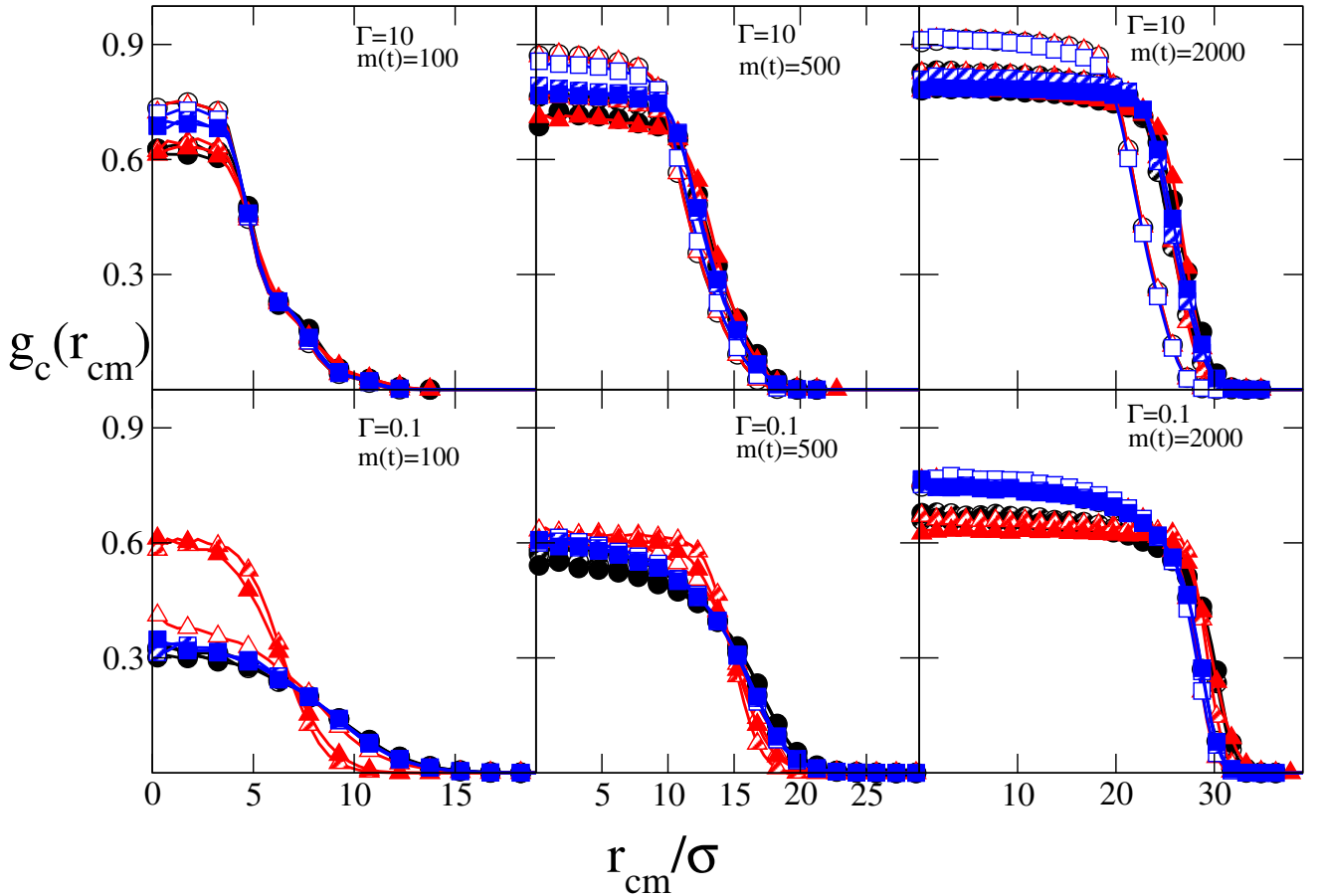}
	\caption{Surface coverage profiles $g(r_{cm})$ for microcolonies with $m(t) = 100, 500$ and $2000$ (left, middle and right column, respectively), and $\Gamma = 10$ and $0.1$ (top and and bottom row, respectively) obtained with SRS model (black lines and circles), attractive Kihara (red lines and triangles) and Hertz model (blue lines and squares). Open, dashed and solid symbols are for $\varepsilon_K= k_BT, 25k_BT$ and $100k_BT$ respectively in the case of SRS and attractive Kihara models, and for and  $\varepsilon_H= 10^4k_BT, 25\cdot10^4k_BT$ and $100\cdot10^4k_BT$ respectively in the case of Hertz model.}
	\label{fig3}
\end{figure*}

\begin{figure}[!h]
	\center
	\includegraphics[width =\columnwidth]{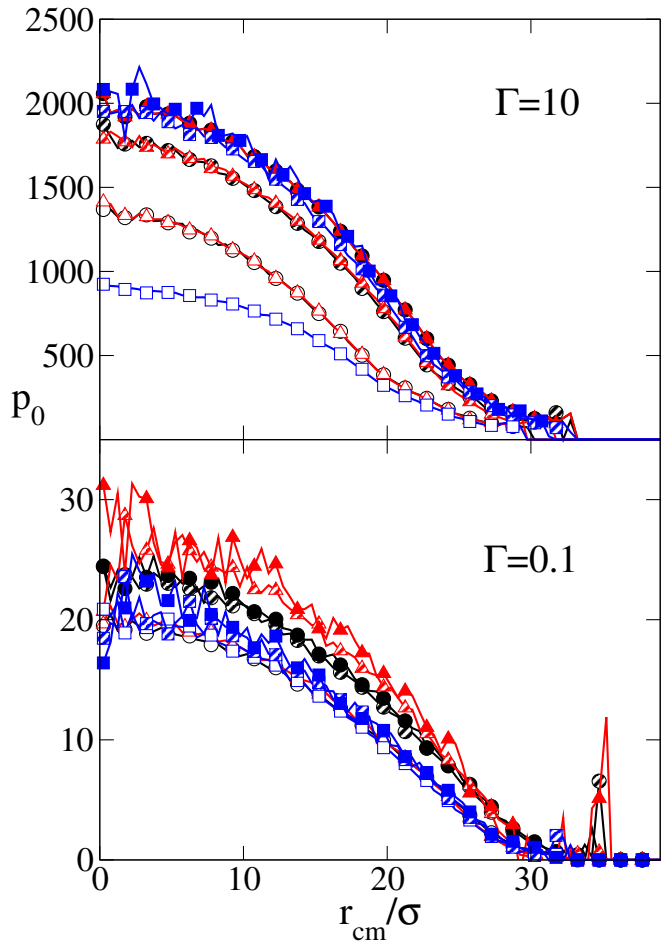}
	\caption{Profile of the isotropic dimensionless pressure $p_0$ felt by bacteria depending on the position respect to the centre of mass of the biofilm $r_{cm}$ at biomass $m(t)=2000$ for $\Gamma=10$ (top panel) and $0.1$ (bottom panel), obtained with SRS model (black lines and circles), attractive Kihara (red lines and triangles) and Hertz model (blue lines and squares). Open, dashed and solid symbols are for $\varepsilon_K= k_BT, 25k_BT$ and $100k_BT$ respectively in the case of SRS and attractive Kihara models, and for  $\varepsilon_H= 10^4k_BT, 25\cdot10^4k_BT$ and $100\cdot10^4k_BT$ in the case of Hertz model.}
	\label{fig4}
\end{figure}

Further insight into this influence can be gleaned from analyzing the behaviour of the coverage profile $g_c(r_{cm})$ as a function of the distance to the center of mass of the microcolony $r_{cm}$. Thus, Fig.\,\ref{fig3} shows this function for the force models and strength explored in this study, for $\Gamma=10$ and $0.1$, at three values of biomass ($m(t)=100, 500, 2000$) representing three stages in the development of microcolonies. The trends we observed in the behavior of $g_c$ can be explained with the same arguments previously used for density, order parameter or eccentricity. Thus, at $\Gamma=10$ (top row of Fig.\,\ref{fig3}), for all the models studied, the microcolony is more compact the lower the strength of the interaction between cells. This effect is clearer as the biomass grows. In contrast, for higher interaction strengths these differences between coverings for different models and strengths tend to disappear.

It is noteworthy that at the lowest biomass value and for $\Gamma=0.1$, the fraction of coverage is higher in the case of potentials with an attractive component than in the rest. This difference is magnified in the case of the attractive Kihara potentials with the high strength  ($\varepsilon_K/k_BT=25$ and $100$), where values for $g_c$ close to $0.6$ are reached in the center of the microcolony, while the rest of the models do not exceed $0.4$. At higher values of biomass and other values of $\Gamma$ these differences do not appear, and the values of $g_c$ for the attractive Kihara models are similar as for the repulsive SRS model with the same value of $\varepsilon_K$. The values of $g_c$ at these values of $m(t)$ and $\Gamma$ are also quite similar to those obtained with the Hertz model. This confirms the relevance of attractive models in situations where cells are dispersed on the surface, forming structures more similar to swarms than to biofilms, but their loss of importance in situations where aggregate structures are formed (high values of $m(t)$ and $\Gamma$). Apart from this specific behavior for the attractive potential, the changes in $g_c$ can also be systematized from the measurement of the inter-particle overlap discussed in Fig.\,\ref{fig1}, with lower coverage in the central zone the higher the inter-particle overlap. At the edge of the microcolony this lower coverage translates into an enlargement of the microcolony size. The latter effect is more evident the higher the biomass value.

Therefore, we have observed how the strength of the forces between bacteria can play an important role in modeling some of the properties of the biofilm. So far we have focused this discussion on the structural properties that have already been discussed in previous works (which focus solely on the SRS model). It will be interesting at this point to analyze the effect on other mechanical properties, which a priory would be more closely related to the force model. For this purpose, we characterize the stress that the particles undergo due to interaction with neighboring cells. To do this, we have calculated the dimensionless stress tensor on particle $i$ as

\begin{equation}\label{eq7}
\displaystyle{p^i_{\alpha,\beta} = \frac{-1}{V_i}\frac{\sigma^3}{k_BT}\sum_{j=1}^{N_c} \textbf{r}^{i_j}_{\alpha}\cdot \textbf{F}^{i_j}_{\beta}}
\end{equation}

where $V_i$ is the volume of bacterium $i$ and the sum is performed for all $N_c$ bacteria $j$ at a distance from bacterium $i$ less than the interaction cut-off. $\textbf{r}^{i_j}_{\alpha}$ is $\alpha$ component of the position vector of the contact point between $i$ and $j$ from the center of cell $i$ and  $\textbf{F}^{i_j}_{\beta}$ is the $\beta$ component of the force exerted by $j$ over $i$. The hydrostatic (and isotropic) pressure that a cell undergoes due to the interaction with the surrounding bacteria can be obtained from the trace of this tensor $p^i_0 =1/2(p^i_{x,x}+p^i_{y,y})$.
From the definition of Eq.\,\ref{eq7}, $p_0$ is defined as a dimensionless quantity. In Fig.\ref{fig4} we show the average of this pressure for all cells at the same distance from the colony's center of mass, $p_0(r_{cm})$, as a function of this distance. Thus, this graph shows the pressure that a cell suffers as a function of its position within the colony. We show in this figure the results for microcolonies in advanced stages, with $m(t)=2000$, for the extreme $\Gamma$ values considered in this study ($\Gamma=10$ in the upper panel and $0.1$ in the lower panel) and for all the models and intensities discussed above.

The behavior of $p_0$ at this value of biomass shows a general dependence on $r_{cm}$, regardless of the value of $\Gamma$. Thus, in both panels of Fig.\,\ref{fig4} it can be seen how $p_0$ shows a high pressure plateau in the center of the microcolony, with a decrease from here towards the edge. In any case, this plateau is narrow, and a general feature is the inhomogeneity in the pressure felt by the cells within the microcolony. This dependence has been reported by other authors in the past for tissues \cite{SHR05} and bacteria \cite{YOU18}. 

Focusing now on the dependence of $p_0$ on $\Gamma$, the comparison of both panels of Fig.\,\ref{fig4} allows us to establish that, the larger $\Gamma$ is, the higher values $p_0$ reaches. Thus, although in the cases shown in the figure the aforementioned trends in the dependence of $p_0$ on the position in the microcolony are maintained, the values obtained for each case show an important variation. Thus, for $\Gamma=10$ the pressure in the center can reach values around $1500$, for $\Gamma=0.1$ it reaches levels between to $20$ and $30$, respectively. These differences in the pressure felt by the cells for each value of $\Gamma$ are related to the compactness of the microcolony, higher the higher the $\Gamma$ due to the lower cell diffusion with respect to growth in this case, and the increase in density (Fig.\,\ref{fig1}) and thus the interaction between bacteria. Surprisingly, the influence of model and force intensity is not as apparent as in previous observable. This is very clear when comparing the SRS model. For $\Gamma=10$ it is found that an increase of $\varepsilon_K=1$ to $100$ only means an increase of about two times in the pressure at the center of the colony (from about $1200$ to $2000$). This change is even smaller for $\Gamma=1$ (from $175$ to $250$, not shown) and almost non-existent for $\Gamma=0.1$. 

Regarding the comparison between models, SRS and attractive Kihara models coincide in the quantitative and qualitative behavior of this magnitude at this amount of biomass for $\Gamma=10$. This was expected given the coincidence of the values of other observables for these two models at high biomass values. Interestingly, for $\Gamma=0.1$ $p_0$ is higher for the attractive Kihara model than for the repulsive one. This is another indication that the attractive interaction might be more relevant in less crowded situations. For the Hertz models of different strength, $p_0$ follows a qualitative and quantitative behavior similar to that found for the SRS model, as can be seen in Fig.\,\ref{fig4}.

\section{Discussion and final remarks}

In summary, in this work we have addressed, through extensive simulations under different conditions, the relevance of the choice of force model when simulating the development of bacterial biofilms using agent-based models. A study like this is relevant because there is not much information on direct experimental measurements of the force between two cells as a function of their shape, size or relative distance. Therefore, the models chosen in the different studies available in the literature are extrapolations of models developed in other fields, such as elastic materials or colloidal systems \cite{GAR12}. In the absence of direct measurements, the realism and goodness of the models used must be validated indirectly through the characteristics of the cell colonies obtained by simulation.

From our results we conclude that, although there are differences when using one model or another, for example SRS, attractive Kihara model, or Hertz in our work, the factor with the most influence on the structural properties of the biofilm is the intensity of the interaction between the cells, rather than how these forces are modeled. Thus, low strengths of the interaction forces favor the appearance of very high overlaps between the cells, which are to be considered unrealistic. When the strength of the forces is increased, these overlaps are reduced and eventually disappear. From a certain threshold of this strength, the results begin to be independent of the parameter that controls the strength of the force, and even of the force model itself that is being applied.

From the results presented here, it is deduced that there is a correlation between the choice of the model for the forces between the cells and their intensity and some of the properties obtained from the growing colonies. These correlations can be explained by the level of overlap between the particles observed for the different cases. For example, the density of the colony is higher the higher the level of overlap between particles, which is relevant for low force intensities. This variation in density translates into other indicators of colony shape such as eccentricity or internal order such as the nematic order parameter, which previous works \cite{ACE18,DEL22} have shown to be strongly dependent on the level of compactness of the colony. But in any case it does not seem to be a particularly relevant effect, and for example a variation of two orders of magnitude in the intensity of the forces causes a much smaller change in these magnitudes, the effect of other parameters such as $\Gamma$ being more relevant. An example of this limited influence is the value of the pressure that the bacteria suffer from being in the center of the colony. One would expect this magnitude to have a direct relationship with the increase in force intensity, however the influence is much milder, with only a doubling of the increase when the intensity increases by two orders of magnitude. However, taking into account the high Young's modulus reported in several works \cite{TUS12} for the interaction between bacteria, and the independence of the results when the intensity of the forces increases, it seems advisable to use these high values for modeling the growth of bacterial colonies. We refer to values higher than $25 k_BT$ for SRS and attractive Kihara models, and higher than $25\cdot10^4k_BT$ for the case of the Hertz model. It should be noted that our model does not introduce a limitation to the growth of bacteria due to the effect of the pressure they suffer, an aspect that has been widely reported in tissues \cite{SHR05}, but more rarely in bacterial colonies \cite{WIT23}. This is probably a general limitation of bacterial growth models, which our group intends to study in future work.

The existence of an attractive contribution to the interaction between bacteria deserves a separate mention. It follows from our results that this possible attractive interaction is only relevant in situations where the colony is not very compact. This means small $\Gamma$ values (situations where diffusion is high and dominant versus cell growth and division), and low biomass values. That is, when cells are initially separated and have not yet filled the space by cell reproduction. In other cases, situations with high $\Gamma$ values where colonies are compact from the early stages, or in more mature colonies with higher biomass, the results found are fully equivalent to those of purely repulsive models. This conclusion follows from the comparison of the attractive and repulsive SRS models. It should be noted that this comparison could be made with other models, for example by comparing the repulsive Hertz model studied in this paper with the Johnson-Kendall-Roberts contact model, which is a modification of the Hertz model by introducing an attractive \cite{DON13,SPO05,ZHA11} contribution. We believe that, in any case, the conclusions would be the same as those found in the comparison of the SRS and attractive Kihara models, and that this additional comparison would not provide qualitatively different results.

Finally, the aim of this work was to study how different models of interaction between bacteria can modify the emergent properties that appear when a bacterial colony develops, with the conclusions that we have summarized in this section. It should be remembered here that our model presents, as an important simplification, the non-explicit consideration of the extracellular matrix, and that we have focused our study on the initial phases of the biofilm, when it is two-dimensional. But we believe that if we were to extend the model by introducing the presence of the extracellular matrix and the possibility of the third dimension, the qualitative results would not change significantly. Of course, introducing these two aspects would significantly modify the evolution of the colony. For example, we have recently done work where, based on our model, the system is three-dimensional, finding an interesting phenomenology \cite{LOB24}. But we understand that the qualitative comparison in this situation between the effects of using different models for cell-cell interaction would not be different. In this sense, we believe that the results of this work can be of general application in the study of bacterial colony development, even in agent-based models that include more ingredients and complexity.


\begin{acknowledgments} 
This work was supported by the Spanish Ministerio de Ciencia e Innovaci\'on and FEDER (Projects n. PID2021-126121NB-I00)./CHU A.R.R. acknowledges financial support from Grant No. PID2021-126348NB-I00 funded by MCIN/AEI/10.13039/501100011033. We thank C3UPO (Universidad Pablo Olavide)for the HPC facilities provided.
\end{acknowledgments}

\bibliography{biblio}


\end{document}